\documentclass[lettersize,journal]{IEEEtran}
\usepackage{amsmath,amsfonts}
\usepackage{algorithmic}
\usepackage{algorithm}
\usepackage{array}
\usepackage[caption=false,font=normalsize,labelfont=sf,textfont=sf]{subfig}
\usepackage{textcomp}
\usepackage{stfloats}
\usepackage{url}
\usepackage{cite}
\usepackage{verbatim}
\usepackage{graphicx}
\usepackage{setspace}
\usepackage{booktabs}
\usepackage{enumerate}
\usepackage{diagbox}
\usepackage{arydshln}
\usepackage{accents}
\usepackage{balance}
\usepackage{bm}
\usepackage{xcolor}

\def\BibTeX{{\rm B\kern-.05em{\sc i\kern-.025em b}\kern-.08em
    T\kern-.1667em\lower.7ex\hbox{E}\kern-.125emX}}
\usepackage{balance}
\begin{document}
\title{{Generalized Code Index Modulation Aided AFDM for Spread Spectrum Systems}}
\author{Mi Qian, \textit{Student~Member,~IEEE}, Fei Ji, \textit{Member,~IEEE}, Yao Ge, \textit{Member,~IEEE}, \\Miaowen Wen, \textit{Senior Member,~IEEE}, and Yong Liang Guan, \textit{Senior Member,~IEEE} 
\thanks{This work was supported in part by the National Natural Science Foundations of China under Grant No.62192712. The work of Yao Ge was supported by the RIE2025 Industry Alignment Fund-Industry Collaboration Projects (IAF-ICP) Funding Initiative, as well as cash and in-kind contribution from the industry partner(s). ({\it{Corresponding authors: Fei Ji; Yao Ge}})}
\thanks{Mi Qian, Fei Ji, and Miaowen Wen are with the School of Electronic and Information Engineering, South China University of Technology, Guangzhou 510641, China (e-mail: eemqian@mail.scut.edu.cn, \{eefeiji, eemwwen\}@scut.edu.cn).}
\thanks{Yao Ge and Yong Liang Guan are with the Continental-NTU Corporate Lab, Nanyang Technological University, Singapore (e-mail: yao.ge@ntu.edu.sg; eylguan@ntu.edu.sg).}}

\maketitle

\begin{abstract}
The recently proposed affine frequency division multiplexing (AFDM) is a new transmission waveform that has shown excellent performance in high-mobility environments, making it a sensible option for the next-generation wireless networks. In this paper, we investigate an energy-efficient generalized code index modulation scheme for AFDM by leveraging spread spectrum, referred to as GCIM-AFDM-SS, to combat the interference caused by the doubly dispersive channels. Specifically, the information bits are conveyed by the transmitted symbols as well as the indices of the selected spreading codes in our proposed GCIM-AFDM-SS scheme. To avoid extensive computations, we also develop a low-complexity maximal ratio combining (MRC) detector algorithm, which recovers the spreading codes first and demodulates the symbols afterwards. Moreover, an upper bound on the bit error rate (BER) of the proposed GCIM-AFDM-SS system with maximum-likelihood (ML) detection is derived. Numerical results demonstrate the superiority of the proposed GCIM-AFDM-SS system over the classical AFDM spread spectrum (AFDM-SS) and the existing index modulated AFDM (IM-AFDM) systems. 
\end{abstract}

\begin{IEEEkeywords}
AFDM modulation, index modulation, spread spectrum, doubly selective fading channels.
\end{IEEEkeywords}

\section{Introduction}
The sixth-generation (6G) wireless networks are expected to achieve ultra-reliable, high energy-efficient, and wide coverage communications in high mobility wireless communications, such as high-speed railway and vehicle-to-everything communications \cite{9779322}. In these scenarios, wireless channels can be characterized by large delay and Doppler spreads, which poses a huge challenge for reliable communications. The widely developed orthogonal frequency division multiplexing (OFDM) systems may not be an optimal solution because of the terrible performance degradation caused by the inter-carrier-interference (ICI) \cite{9380189}. Therefore, developing a new modulation waveform to cope with various challenges in high-mobility communication scenarios is essential.

Against this background, a promising chirp-based multicarrier waveform, named affine frequency division multiplexing (AFDM) was proposed in \cite{10087310,10845819} with high robustness to Doppler spread. The key idea of AFDM is to multiplex the information symbols into a set of orthogonal chirps via inverse discrete affine Fourier transform (IDAFT). By tuning the chirp rate to accommodate the maximum delay spread and maximum Doppler shifts, AFDM can separate all channel paths, ensuring that each symbol experiences multipath fading. Therefore, AFDM can exploit the full diversity in the doubly dispersive channels, achieving performance comparable to that of orthogonal time-frequency space (OTFS) \cite{10891132,10159363} and superior to that of traditional OFDM. Furthermore, AFDM enjoys lower implementation complexity and pilot overhead than OTFS since the two-dimensional delay-Doppler domain in OTFS is transformed into a one-dimensional discrete affine Fourier transform (DAFT) domain, leading to a lower guard pilots \cite{10566604,10557524,10812762}. 

Recently, a novel multicarrier transmission technique known as index modulation (IM) aided AFDM, termed as IM-AFDM, has been introduced to enhance the bit error rate (BER) performance over high-mobility communications \cite{10342712},\cite{10975107}. In \cite{7317808}, an energy-efficient code index modulation (CIM) system has been proposed to improve the data rate without adding extra hardware complexity. By embedding the indices (such as antennas, subcarriers, and time slots) into the transmitted signal, the IM-AFDM scheme significantly enhances both spectral and energy efficiency \cite{8417419},\cite{10381617}. Due to the advantages of IM, several studies have focused on combining AFDM with subcarrier IM, yielding promising results. However, in conventional IM-AFDM schemes, although there is no symbol transmission for the inactive chirp subcarriers, it still experiences severe inter-carrier interference since the fractional Doppler spreads across the entire DAFT domain. To the best of our knowledge, the existing IM-AFDM works have neglected this pivotal problem and exhibit performance loss. 

To alleviate the ICI in doubly dispersive channels, a generalized code index modulation scheme based on AFDM spread spectrum (AFDM-SS), termed as GCIM-AFDM-SS, is proposed in this paper. This innovative study is the first of its kind to exploit the potential of spread spectrum with IM-AFDM to overcome the Doppler spread in the doubly selective fading channels. The main idea is to select an orthogonal code to spread an $M$-ary constellation symbol over the chirp subcarriers, utilizing the indices of spreading code to convey extra information bits. 
The contributions of this paper can be summarized as follows: 1) We propose a novel GCIM-AFDM-SS scheme to effectively overcome the ICI caused by high Doppler spread, ensuring superior BER performance. 2) We also develop a low-complexity maximal ratio combining (MRC) detection algorithm for GCIM-AFDM-SS, which exhibits robustness against the imperfect channel state information (CSI). 3) Numerical results demonstrate that the derived BER upper bound curves match closely with the simulation results, and the proposed GCIM-AFDM-SS scheme significantly outperforms the benchmark schemes in high-mobility channels. 

\begin{figure}
	\centering
	\includegraphics[width=3.4in,height=1.6in]{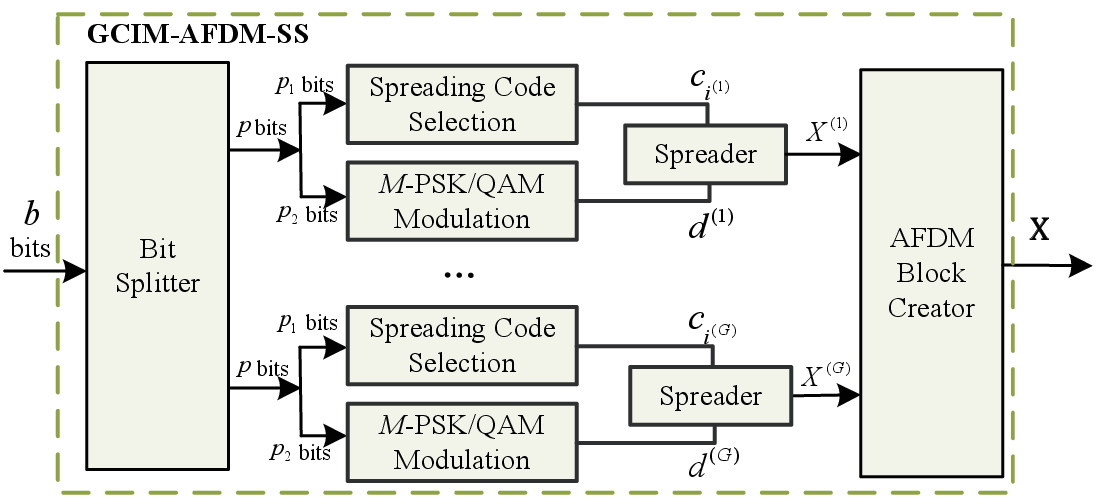}
	\caption{Block diagram of the transmitter structure for the proposed GCIM-AFDM-SS scheme.}
	\label{fig:fig1}
\end{figure}

\section{System Model}
\subsection{Transmitter}
Let $T_s$ denote the sample period, and $N$ denote the the number of chirp subcarriers. The GCIM-AFDM-SS symbols are transmitted with a bandwidth $B=1/T_s$, and subcarrier spacing $\Delta f=B/N=1/{NT_s}$. For each AFDM block, a total of $b$ information bits are transmitted. As shown in Fig.~\ref{fig:fig1}, the $b$ bits are separated into $G$ subblocks, where each subblock containing $p=b/G$ bits. Since the IM mapping operation within $n=N/G$ subcarriers in each subblock is the same and independent, let us take the $g$-th subblock as an example, where $g=\{1,\dots ,G\}$. The transmitted bits of our proposed GCIM-AFDM-SS system are carried by the index of the selected spreading code as well as the $M$-ary modulated symbols. In the $g$-th subblock, the incoming $p$ bits are first divided into two parts such that $p=p_1+p_2$, where $p_1=\log_{2}{n}$ for selecting a Walsh-Hadamard spreading code, and $p_2=\log_{2}{M}$ for choosing an $M$-ary constellation symbol, respectively. The first part of $p_1$ bits are applied to the index mapper to determine the activated spreading codes $\mathbf{c}_{{i}^{(g)}}\in \mathbb{C}^{n \times 1}$ from a predefined set $\mathcal{C} = \{\mathbf{c}_{1},\dots,\mathbf{c}_{n}\}$, where ${i}^{(g)} \in\{1,\dots,n\}$ denotes the index of the spreading code for the $g$-th subblock. The second part is applied to the modulation symbol index mapper to determine the activated symbols $d^{(g)}\in \mathcal{X}$, where $\mathcal{X}=\{a_1,\dots,a_t,\dots,a_M\}$ with $t\in[1,M]$ is the $M$-ary phase shift keying (PSK) or quadrature amplitude modulation (QAM) constellation set. The SE of the proposed GCIM-AFDM-SS system is calculated as $R=(\log_{2} {n}  +\log_{2}{M})G/N$.

During the GCIM-AFDM-SS process, the transmitted symbol of the $g$-th subblock $d^{(g)}$ is spreaded across $n$ subcarriers by the selected spreading code $\mathbf{c}_{{i}^{(g)}}$, yielding
\begin{align}
\textbf{x}^{(g)}&=d^{(g)}\mathbf{c}_{i^{(g)}}=[x_{1}^{(g)},\dots,x_{n}^{(g)}]^T\nonumber\\
&= [d^{(g)}c_{i^{(g)},1},\dots,d^{(g)}c_{i^{(g)},n}]^T,
\end{align}
where $c_{i^{(g)},k}$, $k\in\{1,\dots,n\}$ denotes the $k$-th element of $\mathbf{c}_{{i}^{(g)}}$.

After obtaining the index information and the constellation symbol of all the GCIM-AFDM-SS subblocks, the AFDM block creator generates a discrete affine Fourier domain symbol $\textbf{x}\in \mathbb{C}^{N\times 1}$ as   
\begin{equation}
\textbf{x} = [\textbf{x}^{(1)},\dots,\textbf{x}^{(G)}]^T.
\end{equation}

The corresponding AFDM time domain signal $\mathbf{s}$ is given by inverse discrete affine Fourier transform (IDAFT) on $\textbf{x}$, i.e.,
\begin{align}
s[n]=\frac{1}{\sqrt N}\sum_{m=0}^{N-1}&x[m]e^{j2\pi(c_{1}n^2+c_{2}m^2+nm/N)}, \nonumber\\ 
&n=0,\dots,N-1,
\label{eq:sn}
\end{align}
where the adjustable DAFT parameters $c_1\geq 0$ and $c_2\geq 0$ are optimized to prevent different channel delay and Doppler overlapping. Note that (\ref{eq:sn}) can be written in a matrix form as
\begin{align}
\textbf{s}=\mathbf{A}^{H}\textbf{x}=\mathbf{\Lambda} _{c_{1}}^{H}\textbf{F}^{H}\mathbf{\Lambda}_{c_{2}}^{H}\textbf{x},
\end{align}
where $\mathbf{A}=\mathbf{\Lambda} _{c_{2}}\textbf{F}\mathbf{\Lambda}_{c_{1}}$ denotes the DAFT matrix, $\textbf{F}$ represents the discrete Fourier transform (DFT) matrix with entries $e^{-j2\pi mn/N}/\sqrt{N}$ and $\mathbf{\Lambda} _{c_i}=\textup{diag}(e^{-j2\pi c_in^2},n=0,\dots,N-1,i=1,2)$. 

Before transmitting the time-domain signal $\mathbf{ s }$, a prefix is required to combat multipath propagation and makes the channel lie in a periodic domain. Unlike OFDM, AFDM employs a chirp-periodic prefix (CPP) here instead of a cyclic prefix (CP) due to the inherent periodicity of DAFT matrix. We assume that the length of CPP  is greater than the maximum path delay spread, to avoid the inter-block interference. 

\subsection{Channel Model}
We consider a linear time-varying (LTV) channel between the transmitter and receiver with $L$ resolvable propagation paths, which can be modeled as\cite{10087310}
\begin{align}
h(\tau ,t )=\sum_{l=1}^{L}h_{l}\delta \left ( \tau -\tau _{l} \right )e^{-j\frac{2\pi}{N}\nu_{l}t},
\end{align} 
where $\delta \left ( \cdot  \right )$ denotes the dirac delta function. The complex channel gain of the $l$-th path $h_l$ follows a distribution of $\mathcal{CN}(0,1/L)$. Each path includes a unique delay shift $\tau _{l} \in [0,\tau_{max}]$ and Doppler shift $\nu _{l}=\alpha _{l}+\beta _{l}\in [-\nu_{max},\nu_{max}]$, where $\tau_{max}$ and $\nu_{max}$ denote the maximum delay shift and maximum Doppler shift, respectively. Here, $\nu_l$ is the Doppler shift normalized with respect to the subcarrier spacing, where $\alpha _{l}\in [-\alpha_{max},\alpha_{max}]$ and $\beta _{l}\in(-\frac{1}{2},\frac{1}{2}]$ denote the integer and fractional components of $\nu_l$, respectively. Note that our channel model is a generalized model that allows each delay tap to be equipped with different Doppler shifts for different paths $l,t\in \{1,\dots,L\}$, while $\tau_l=\tau_t$ and $\nu_l \neq \nu_t$.

\subsection{Receiver}
After transmission over the doubly-selective channel, the received time domain signal can be written as
\begin{align}
r[n] = \sum_{l=1}^{L}h_{l}s[n-l]e^{-j\frac{2\pi }{N} \nu _{l}n}+w[n],
\label{eq:rn}
\end{align} 
where $w\sim \mathcal{CN}(0,N_0)$ represents the complex additive white Gaussian noise (AWGN). 

After discarding the CPP, (\ref{eq:rn}) can be rewritten in a matrix form as follows
\begin{align}
\textbf{r}=\textbf{H}\textbf{s} + \textbf{w},
\end{align} 
where $\textbf{w} \sim \mathcal{CN}(0,N_0\textbf{I})$ is the $N\times 1$ time domain noise vector. $\textbf{H}=\sum_{l=1}^{L}h_{l} \mathbf{\Gamma }_{\textup{CPP}_{l}}\mathbf{\Delta} _{\nu _{l}}\mathbf{\Pi} ^{\tau _{l}}\in \mathbb{C}^{N\times N}$ denotes the effective time domain channel matrix, whose $l$-th path is fully composed of four parameters: a) a complex channel fading coefficient $h_l$, b) a diagonal matrix related to CPP $ \mathbf{\Gamma }_{\textup{CPP}_{l}}\in \mathbb{C}^{N\times N}$, c) a Doppler shift matrix $\mathbf{\Delta} _{\nu _{l}}=\textup{diag}(e^{-j\frac{2\pi }{N}\nu_{l}n},n=0,\dots,N-1)\in \mathbb{C}^{N\times N}$, and d) a forward cyclic-shift matrix $\mathbf{\Pi} ^{\tau _{l}}\in \mathbb{C}^{N\times N}$ \cite{10087310}.

Finally, after performing the $N$-point DAFT on the received signal $\textbf{r}$, one can obtain the output symbols in the DAFT-domain as given by
\begin{align}
y[m]=\frac{1}{\sqrt N}\sum_{n=0}^{N-1}&r[n]e^{-j2\pi(c_{1}n^2+c_{2}m^2+nm/N)} , \nonumber\\ 
&m=0,\dots,N-1.
\label{eq:ym}
\end{align}
%By substituting (\ref{eq:sn}) and (\ref{eq:rn}) into (\ref{eq:ym}), the input-output relationship in the time domain can be calculated as
%\begin{align}
%	y[m]=\frac{1}{\sqrt N}\sum_{n=0}^{N-1}&r[n]e^{-j2\pi(c_{1}n^2+c_{2}m^2+nm/N)} , \nonumber\\ 
%	&m=0,\dots,N-1.
%	\label{eq:IO}
%\end{align}
%We set the DAFT parameter $ c_{1}=\frac{2\left ( \alpha _{max}+1 \right )+1}{2N}$.

The matrix representation of the input-output relationship in the DAFT domain can be written as
\begin{align}
\textbf{y}&=\textbf{A}\textbf{r}= \sum_{l=1}^{L}h_{l}\underset{\textbf{H}_l}{\underbrace{\textbf{A}\mathbf{\Gamma }_{\textup{CPP}_l}\mathbf{\Delta }_{\nu_{l}}\mathbf{\Pi }^{\tau _{l}}\textbf{A}^{H}}}\textbf{x}+\textbf{A}\textbf{w}\nonumber\\ &=\textbf{H}_\textup{{eff}}\textbf{x}+\tilde{\textbf{w}},
\label{eq:yHx}
\end{align} 
where $\textbf{H}_l={\textbf{A}\mathbf{\Gamma }_{\textup{CPP}_l}\mathbf{\Delta }_{\nu_{l}}\mathbf{\Pi }^{\tau _{l}}\textbf{A}^{H}}$ is the $l$-th path subchannel matrix in the DAFT domain, $\textbf{H}_\textup{{eff}}=\sum_{l=1}^{L}h_{l}\textbf{H}_l$ denotes the effective channel matrix, and $\tilde{\textbf{w}}$ represents the effective noise vector that has the same statistical properties as $\textbf{w}$.

Upon receiving the DAFT domain signal $\textbf{y}$, we introduce two detection methods for the proposed GCIM-AFDM-SS scheme, including the ML detector and low-complexity MRC detector. 
%\begin{enumerate}
%\item{ML Detector}: 

 \textbf{1) ML detector:} At the receiver, we need to detect the indices of the active spreading code and the corresponding data symbols. From (\ref{eq:yHx}), the ML detector makes a joint decision by searching all possible index combinations and signal constellation points, which is given by
\begin{align}
(\hat{\textbf{i}},\hat{\textbf{d}})= \underset{\forall \textbf{i},\textbf{d}}{\textup{arg\:min}} \left\| \textbf{y}-\textbf{H}_{\textup{eff}}\textbf{x}\right\|^{2},
\end{align} 
where $\textbf{i}$ and $\textbf{d}$ denote the indices of the selected spreading code and the transmitted $M$-ary modulation symbols, respectively. Here, we denote $\textbf{i}=\left\{ i^{(1)},\dots,i^{(G)}\right\}$ with $i^{(g)} \in \{1,\dots,n\}$, $\textbf{d}=\left\{ d^{(1)},\dots,d^{(G)}\right\}$, and let $d^{(g)}$ represents an $M$-PSK/QAM symbol. The corresponding $b$ transmitted bits can be recovered from $\hat{\textbf{i}}$ and $\hat{\textbf{d}}$. Since $i^{(g)}$ and $d^{(g)}$ possess $2^n$ and $M$ possibilities in each subblock, the computational complexity of the ML detector is easy to be calculated by $\mathcal{O}((2^nM)^{\frac{N}{n}})$. Obviously, the decoding complexity increases exponentially as the increase of $N$, which limits the application of ML to the large values of $N$. To avoid this computation burden, we propose a novel low-complexity detector as below. 

%\item{Low complexity MRC Detector}: 

\textbf{2) Low-complexity MRC detector:} In this detector, the detection process is divided into two steps. The first step is to estimate the activated spreading code, followed by detecting the constellation symbol using the estimated spread code obtained in the first step. 

At the first stage, the received signal $\textbf{y}$ is fed into the MMSE equalizer. Then, the equalized signal $\bar{\textbf{y}}=\mathbf{H}_\textup{{eff}}^{\ast }\textbf{y}$ is divided into $G$ groups, with all the subcarriers in each group undergoing multiplication by a $n\times n$ Walsh matrix for despreading. According to the MRC detector, the output of $g$-th group is given by $\Delta_{g } =\sum_{k=1}^{n}c_{i^{(g)} ,k}^{*}\bar{y}_k,g =1,\dots,G$, where $\bar{y}_k$ denotes $k$-th elements of the equalized signal  $\bar{\textbf{y}}$. To effectively identify the indices of the selected spreading code in each group, we compare the squared values of the outputs of $n$ branches in each group via 
\begin{align}
\hat{i}^{(g)}=\underset{\forall g}{\textup{arg\:max}}\left | \Delta_{g } \right |^{2}.
\end{align}

According to the obtained $\hat{\textbf{i}}=[\hat{i}^{(1)},\dots,\hat{i}^{(g)}]$, in the second stage, we can estimate the constellation symbols by searching all possible signal constellation points, as given by
\begin{align}
\hat{\mathbf{d}}=\underset{\mathbf{d}}{\textup{arg min}}\left \| \textbf{y} -\mathbf{{H}_{eff}}\mathbf{x} \right \|^{2}.
\end{align}

Based on the above analysis, it is obvious that the computation complexity of the proposed GCIM-AFDM-SS detector is of order $\mathcal{O}(N+M^G)$, which is lower than that of the ML algorithm. Therefore, the proposed low-complexity MRC detector is particularly suitable for high-dimensional detection. 

\section{Performance Analysis}
In this section, we derive an upper bound of the average BER of our proposed GCIM-AFDM-SS system under Rayleigh fading channels.

Based on the transmission model (\ref{eq:yHx}), we can represent the DAFT domain signal as 
\begin{align}
\textbf{y}=\sum_{l=1}^{L}h_{l}\textbf{H}_{l}\textbf{x}+\tilde{\textbf{w}}=\mathbf{\Upsilon }( \textbf{x})\textbf{h}+\tilde{\textbf{w}},
\label{eq:YUph}
\end{align} 
where channel gain $\textbf{h}\in \mathbb{C}^{L\times 1}$ is given by $[h_1,\dots,h_L]^T$, and $\mathbf{\Upsilon }(\textbf{x})$ is the $N\times L$ concatenated matrix constructed as follows
\begin{align}
\mathbf{\Upsilon }(\textbf{x})=[\textbf{H}_1\textbf{x}|\dots|\textbf{H}_L\textbf{x}].
\label{eq:Upsilon}
\end{align} 

The elements of $\textbf{x}$ is normalized so that the average energy of $\textbf{x}$ is unit. Assume the perfect channel state information is known at the receiver. The erroneously decoded symbols  $\hat{\textbf{x}}$ can be caused by either indices of active spreading code or constellation symbols when $\textbf{x}$ is transmitted, where $\textbf{x} \neq \hat{\textbf{x}}$. From (\ref{eq:YUph}) and (\ref{eq:Upsilon}), the conditional pairwise error probability (PEP) between two distinct data $\textbf{x}$ and $\hat{\textbf{x}}$ of the proposed GCIM-AFDM-SS can be calculated as
\begin{equation}
P(\textbf{x} \rightarrow \hat{\textbf{x}} | \mathbf{h} )=Q\left(\sqrt{\frac{\mathbf{\Theta }}{2 N_{0}}}\right),
\label{eq:Pxh}
\end{equation}
where $\mathbf{\Theta }=\left\| \left(\mathbf{\Upsilon } (\textbf{x} )-\mathbf{\Upsilon }(\hat{\textbf{x}})\right)\textbf{h}\right\|^2=\textbf{h}^{H}\mathbf{\Gamma }\textbf{h}$ and $\mathbf{\Gamma }=\left(\mathbf{\Upsilon } (\textbf{x} )-\mathbf{\Upsilon }(\hat{\textbf{x}})\right)^H\left(\mathbf{\Upsilon } (\textbf{x} )-\mathbf{\Upsilon }(\hat{\textbf{x}})\right)$. Here, $Q(\cdot )$ represents the tail distribution function of a standard normal distribution. By applying the approximation $Q(x)\approx \frac{1}{12}\textup{exp}(-x^{2}/2)+\frac{1}{4}\textup{exp}(-2x^2/3) $ into (\ref{eq:Pxh}), we can derive an approximate unconditional PEP as 
\begin{figure}
	\center
	\includegraphics[width=3.2in,height=2.12in]{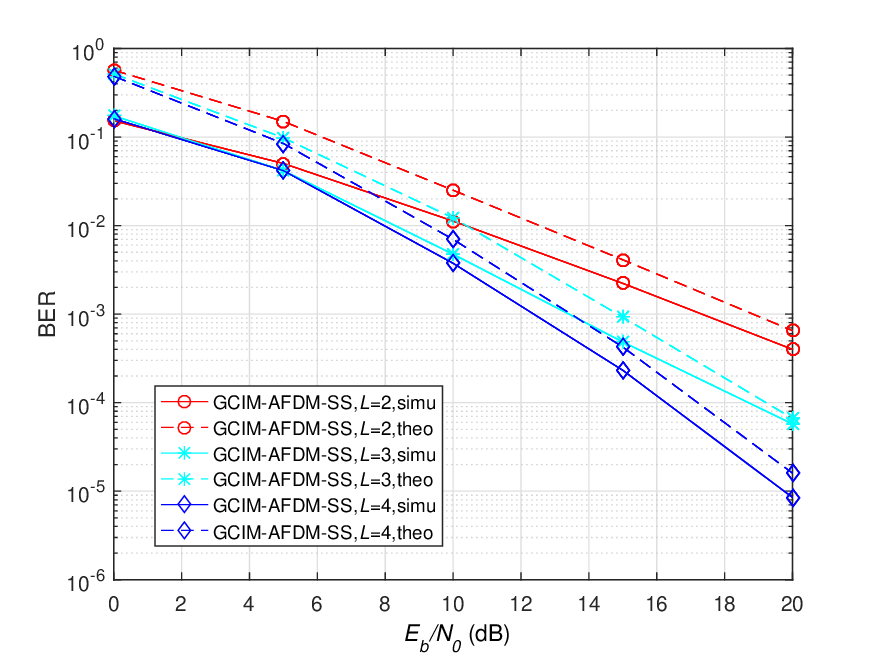}
	\caption{Simulated and analytical BER performance comparisons of the proposed GCIM-AFDM-SS scheme for different numbers of paths.}
	\label{fig:simu_theo}
\end{figure}
\begin{equation}
P(\textbf{x} \rightarrow \hat{\textbf{x}}  )=E_\mathbf{h}\left\{\frac{1}{12}\textup{exp}(- \lambda_{1}\mathbf{\Theta } )+\frac{1}{4}\textup{exp}(-\lambda_2\mathbf{\Theta } )\right\},
\label{eq:Pxhatx}
\end{equation}
where $\lambda_{1}=1/4N_0$ and $\lambda_{2}=1/3N_0$. Since the matrix $\mathbf{\Gamma }$ is a Hermitian matrix, we define its rank and the distinct non-zero eigenvalues as $R$ and $\zeta_q$, $q=1,\dots,R$, respectively. As a results, it follows that
\begin{align}
\left\| \left(\mathbf{\Upsilon } (\textbf{x} )\!-\!\mathbf{\Upsilon }(\hat{\textbf{x}})\right)\textbf{h}\right\|^2&=\textbf{h}^H\left(\mathbf{\Upsilon } (\textbf{x} )\!-\!\mathbf{\Upsilon }(\hat{\textbf{x}})\right)^H\left(\mathbf{\Upsilon } (\textbf{x} )\!-\!\mathbf{\Upsilon }(\hat{\textbf{x}})\right)\textbf{h}
\nonumber\\
&=\textbf{h}^H \textbf{U}\mathbf{\Lambda }\textbf{U}^{H}\textbf{h}\nonumber
=\vec{\textbf{h}}^H\mathbf{\Lambda }\vec{\textbf{h}}\nonumber\\
&=\sum_{q=1}^{R}\zeta _{q}\left | \vec{h}_{q}\right |^{2},
\label{eq:Gamma_H}
\end{align}
where $\textbf{U}$ denotes a unitary matrix and $\mathbf{\Lambda }=\textup{diag}\left\{ \zeta _{1},\dots,\zeta _{R}\right\}$ is a diagonal matrix. In (\ref{eq:Gamma_H}), $\vec{h}_{q}$ represents the $q$-th element of the vector $\vec{\textbf{h}}=\textbf{U}^H\textbf{h}$, which is obtained by multiplying $\textbf{h}$ with a unitary matrix. Thus, $\vec{\textbf{h}}$ has the same distribution as $\textbf{h}$. 

Assuming each path has independent Rayleigh distribution, the unconditional PEP in (\ref{eq:Pxhatx}) can be approximated as
\begin{equation}
P(\textbf{x} \rightarrow \hat{\textbf{x}})\cong \frac{1}{12}\prod_{q=1}^{R}\frac{1}{1+\frac{\lambda _{1}\zeta _{q}}{L}}+\frac{1}{4}\prod_{q=1}^{R}\frac{1}{1+\frac{\lambda _{2}\zeta _{q}}{L}}. 
\label{eq:PEPapprox}
\end{equation}

At high SNRs, i.e., $1/N_0\to +{\infty}$, (\ref{eq:PEPapprox}) can be further approximated to
\begin{equation}
P(\textbf{x} \rightarrow \hat{\textbf{x}}) \approx N_{0}^{-R}\left ( \frac{4^{R}}{12} +\frac{3^{R}}{4}\right )\prod_{q=1}^{R}\zeta_{q}^{-1}.
\end{equation}

According to the union bounding technique, we can derive the average bit error probability (ABEP) upper bound of the proposed GCIM-AFDM-SS system as
\begin{equation}
P_{\textup{ABEP}}\leq \frac{1}{b2^{b}}\sum_{\textbf{x}}\sum_{\hat{\textbf{x}}}P(\textbf{x} \rightarrow \hat{\textbf{x}})e(\textbf{x} \rightarrow \hat{\textbf{x}}),
\label{eq:ABEP}
\end{equation}
where $e(\textbf{x} \rightarrow \hat{\textbf{x}})$ denotes the Hamming distance for an arbitrary pair of $(\textbf{x},\hat{\textbf{x}})$ with $\textbf{x}\neq \hat{\textbf{x}}$.

\begin{figure}
	\center
	\includegraphics[width=3.2in,height=2.12in]{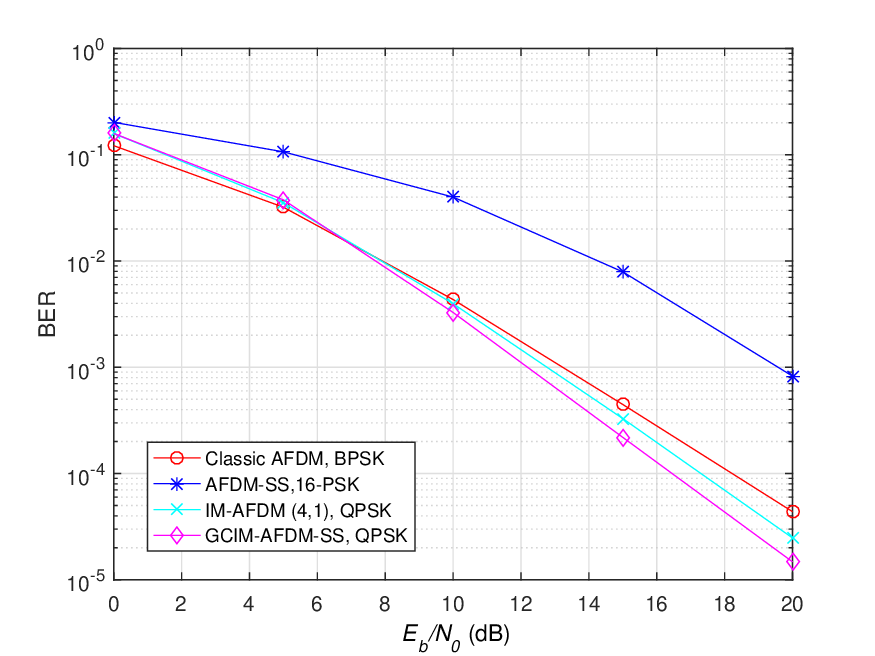}
	\caption{BER performance comparison between the proposed GCIM-AFDM-SS, classic AFDM, AFDM-SS and IM-AFDM schemes with ML detector at the same SE (i.e., 1 bps/Hz), where $N=8, n=4, L=3$.}
	\label{fig:ML_compare}
\end{figure}

\section{Numerical Results}
In this section, we present the BER performance of  the proposed GCIM-AFDM-SS system with chirp subcarrier spacing of 15 kHZ and carrier frequency of 4 GHz. The maximum Doppler shift is set as $\alpha_{max}=1$, which corresponds to a maximum speed of 405km/h. Besides, we consider $ c_{1}=\frac{2\left ( \alpha _{max}+1 \right )+1}{2N}$ and $c_2$ is set as an arbitrary irrational number. Each path is assumed to be independent and the channel gains of all paths follow the distribution of $\mathcal{CN}(0,1/L)$. Unless otherwise stated, the fractional Doppler case is employed, and the fractional Doppler shift of each path is generated by Jake's formula \cite{10087310}. Notably, ``IM-AFDM $(n,n')$" is referred to as the IM-AFDM scheme in which $n'$ out of $n$ chirp subcarriers are active. For comparison, we set the classical AFDM \cite{10087310}, IM-AFDM \cite{10975107} and AFDM-SS schemes as the benchmarks.

Fig.~\ref{fig:simu_theo} illustrates the analytical and simulated BER performance of the ML detector for the proposed GCIM-AFDM-SS scheme, where $N=4$ and QPSK are employed. One can observe that all the simulated curves closely match with the analytical BERs in the high SNR region, which confirms the effectiveness of our PEP theoretical analysis. A discrepancy between the analytical and simulated BER is observed at low SNRs as the approximation in (\ref{eq:ABEP}) is more precise at high SNRs. Moreover, the proposed GCIM-AFDM-SS scheme enjoys superior performance improvement as the number of channel paths increases from $L=2$ to $L=4$. This is due to the fact that a larger diversity gain can be exploited as the number of independent resolvable paths increases.

In Fig.~\ref{fig:ML_compare}, we compare the BER performance of GCIM-AFDM-SS with QPSK, classical AFDM with BPSK, AFDM-SS with 16-PSK, and IM-AFDM with $(n,n')=(4,1)$ and QPSK at the same SE of 1 bps/Hz. One can notice that the proposed GCIM-AFDM-SS scheme outperforms both the classical AFDM, AFDM-SS and IM-AFDM schemes in the medium-to-high SNR regime. Approximately 1dB gain can be achieved at a BER level of $10^{-4}$ by GCIM-AFDM-SS scheme compared with the conventional IM-AFDM scheme. The performance gain of our GCIM-AFDM-SS scheme can be interpreted by the fact that the spreading code achieves superior inter-carrier interference mitigation compared to partial subcarrier data transmission. The AFDM-SS system demonstrates limited performance as higher-order modulation results in severe performance deterioration.

\begin{figure}
	\center
	\includegraphics[width=3.2in,height=2.12in]{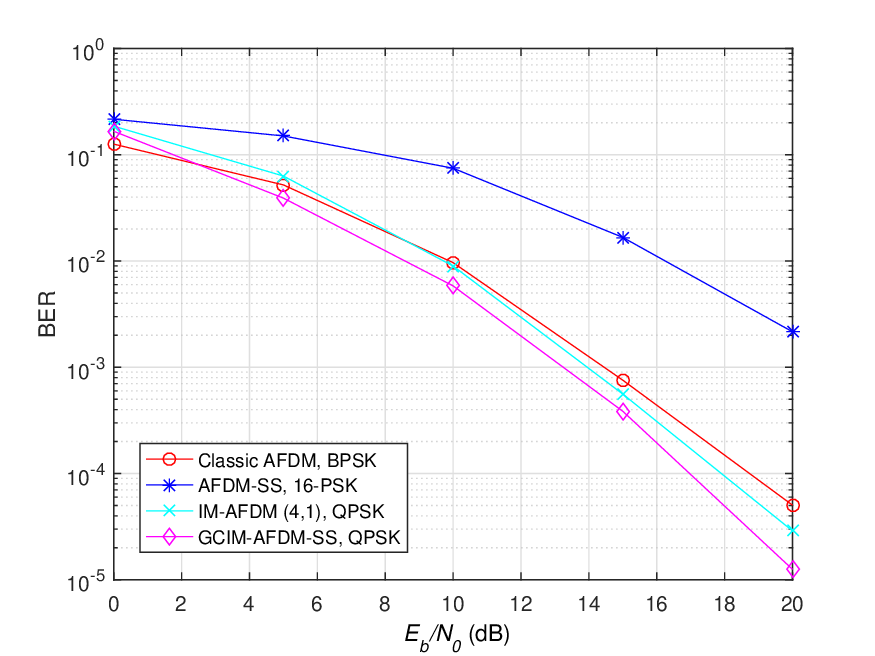}
	\caption{BER performance comparison between the proposed GCIM-AFDM-SS scheme, the classical AFDM, AFDM-SS and IM-AFDM with low-complexity MRC detector under the same SE (i.e., 1 bps/Hz), where $N=64, n=4, L=14$.}
	\label{fig:MMSE_compare}
\end{figure}

Fig.~\ref{fig:MMSE_compare} compares the BER performance of the GCIM-AFDM-SS scheme with QPSK, and that of the classical AFDM with BPSK, AFDM-SS with 16-PSK, and IM-AFDM with $(n,n')=(4,1)$ and QPSK by using the proposed low-complexity MRC detector. Similar to the observations in Fig.~\ref{fig:ML_compare}, the proposed GCIM-AFDM-SS scheme achieves a higher performance gain than the conventional benchmark schemes. These results powerfully identify the superiority and dependability of our proposed GCIM-AFDM-SS scheme and MRC detector in terms of high-dimensional transmissions.

\begin{figure}
	\center
	\includegraphics[width=3.2in,height=2.12in]{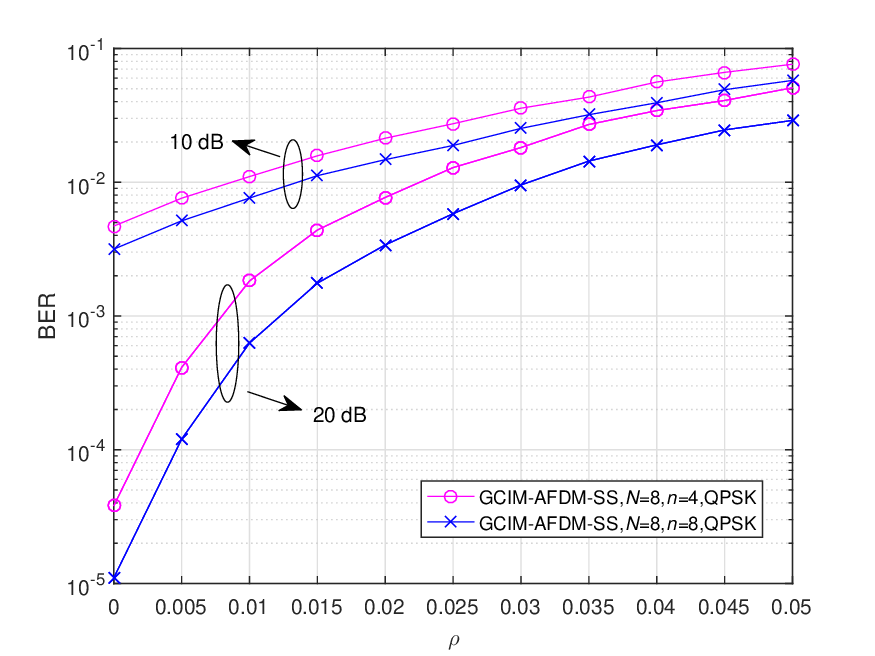}
	\caption{BER performance of the proposed GCIM-AFDM-SS scheme under different lengths of spreading code with imperfect CSI.}
	\label{fig:rho_BER}
\end{figure}

In Fig.~\ref{fig:rho_BER}, the performance of our proposed GCIM-AFDM-SS scheme with different length of spreading code is tested for imperfect CSI. As shown in Fig.~\ref{fig:rho_BER}, our analysis indicates that the performance loss of the proposed GCIM-AFDM-SS scheme remains mild even under moderate values of channel uncertainty $\rho$. The acceptable degradation of BER performance demonstrates the robustness of our proposed GCIM-AFDM-SS scheme against channel modeling errors. Moreover, a higher coding gain can be obtained by increasing the length of spreading code. The proposed GCIM-AFDM-SS scheme provides a flexible trade-off between the SE and BER performance by modifying the length of spreading code $n$.

\section{Conclusion}
In this paper, we proposed an innovative IM-AFDM structure with spread spectrum, termed as GCIM-AFDM-SS, to overcome the ICI of the doubly dispersive channels under fractional delay/Doppler shifts. The theoretical BER upper bound for the proposed GCIM-AFDM-SS scheme was derived and validated by the Monte Carlo simulations. The performance comparison results of both the ML and low-complexity MRC detectors indicated that the proposed GCIM-AFDM-SS scheme can achieve at least 1 dB gain with respect to the benchmark schemes. In the future, we will explore our proposed scheme with multiple-mode IM, multiple access, and multiple-input-multiple-output (MIMO) systems. 

\bibliographystyle{IEEEtran}
\bibliography{reference}

\begin{thebibliography}{10}
\providecommand{\url}[1]{#1}
\csname url@samestyle\endcsname
\providecommand{\newblock}{\relax}
\providecommand{\bibinfo}[2]{#2}
\providecommand{\BIBentrySTDinterwordspacing}{\spaceskip=0pt\relax}
\providecommand{\BIBentryALTinterwordstretchfactor}{4}
\providecommand{\BIBentryALTinterwordspacing}{\spaceskip=\fontdimen2\font plus
\BIBentryALTinterwordstretchfactor\fontdimen3\font minus
  \fontdimen4\font\relax}
\providecommand{\BIBforeignlanguage}[2]{{%
\expandafter\ifx\csname l@#1\endcsname\relax
\typeout{** WARNING: IEEEtran.bst: No hyphenation pattern has been}%
\typeout{** loaded for the language `#1'. Using the pattern for}%
\typeout{** the default language instead.}%
\else
\language=\csname l@#1\endcsname
\fi
#2}}
\providecommand{\BIBdecl}{\relax}
\BIBdecl

\bibitem{9779322}
M.~{Noor-A-Rahim}, Z.~{Liu}, H.~{Lee}, M.~O. {Khyam}, J.~{He}, D.~{Pesch},
  K.~{Moessner}, W.~{Saad}, and H.~V. {Poor}, ``{6G for vehicle-to-everything
  (V2X) communications: Enabling technologies, challenges, and
  opportunities},'' \emph{Proc. IEEE}, vol. 110, no.~6, pp. 712--734, Jun.
  2022.

\bibitem{9380189}
M.~{Wen}, J.~{Li}, S.~{Dang}, Q.~{Li}, S.~{Mumtaz}, and H.~{Arslan},
  ``{Joint-mapping orthogonal frequency division multiplexing with subcarrier
  number modulation},'' \emph{IEEE Trans. Commun.}, vol.~69, no.~7, pp.
  4306--4318, Jul. 2021.

\bibitem{10087310}
A.~Bemani, N.~Ksairi, and M.~Kountouris, ``{Affine frequency division
  multiplexing for next generation wireless communications},'' \emph{IEEE
  Trans. Wireless Commun.}, vol.~22, no.~11, pp. 8214--8229, Nov. 2023.

\bibitem{10845819}
Y.~Tao, M.~Wen, Y.~Ge, J.~Li, E.~Basar, and N.~Al-Dhahir, ``{Affine frequency
  division multiplexing with index modulation: Full diversity condition,
  performance analysis, and low-complexity detection},'' \emph{IEEE J. Sel.
  Areas Commun.}, vol.~43, no.~4, pp. 1041--1055, Apr. 2025.

\bibitem{10891132}
\BIBentryALTinterwordspacing
Q.~Deng, Y.~Ge, and Z.~Ding, ``{A unifying view of OTFS and its many
  variants},'' \emph{IEEE Commun. Surv. Tuts.}, pp. 1--1, 2025. [Online].
  Available:
  \url{https://ieeexplore.ieee.org/stamp/stamp.jsp?arnumber=10891132}
\BIBentrySTDinterwordspacing

\bibitem{10159363}
M.~Qian, F.~Ji, Y.~Ge, M.~Wen, X.~Cheng, and H.~V. Poor, ``{Block-wise index
  modulation and receiver design for high-mobility OTFS communications},''
  \emph{IEEE Trans. Commun.}, vol.~71, no.~10, pp. 5726--5739, Oct. 2023.

\bibitem{10566604}
Q.~Luo, P.~Xiao, Z.~Liu, Z.~Wan, N.~Thomos, Z.~Gao, and Z.~He, ``{AFDM-SCMA: A
  promising waveform for massive connectivity over high mobility channels},''
  \emph{IEEE Trans. Wireless Commun.}, vol.~23, no.~10, pp. 14\,421--14\,436,
  Oct. 2024.

\bibitem{10557524}
H.~Yin, X.~Wei, Y.~Tang, and K.~Yang, ``{Diagonally reconstructed channel
  estimation for MIMO-AFDM with inter-doppler interference in doubly selective
  channels},'' \emph{IEEE Trans. Wireless Commun.}, vol.~23, no.~10, pp.
  14\,066--14\,079, Oct. 2024.

\bibitem{10812762}
H.~Yuan, Y.~Xu, X.~Guo, Y.~Ge, T.~Ma, H.~Li, D.~He, and W.~Zhang, ``{PAPR
  reduction with pre-chirp selection for affine frequency division
  multiplexing},'' \emph{IEEE Wireless Commun. Lett.}, vol.~14, no.~3, pp.
  736--740, Mar. 2025.

\bibitem{10342712}
J.~Zhu, Q.~Luo, G.~Chen, P.~Xiao, and L.~Xiao, ``{Design and performance
  analysis of index modulation empowered AFDM system},'' \emph{IEEE Wireless
  Commun. Lett.}, vol.~13, no.~3, pp. 686--690, Mar. 2024.

\bibitem{10975107}
\BIBentryALTinterwordspacing
G.~Liu \emph{et~al.}, ``{Pre-chirp-domain index modulation for full-diversity
  affine frequency division multiplexing towards 6G},'' \emph{IEEE Trans.
  Wireless Commun.}, pp. 1--1, 2025. [Online]. Available:
  \url{https://ieeexplore.ieee.org/stamp/stamp.jsp?tp=&arnumber=10975107}
\BIBentrySTDinterwordspacing

\bibitem{7317808}
G.~Kaddoum, Y.~Nijsure, and T.~Hung, ``{Generalized code index modulation
  technique for high-data-rate communication systems},'' \emph{IEEE Trans. Veh.
  Technol.}, vol.~65, no.~9, pp. 7000--7009, Sep. 2016.

\bibitem{8417419}
T.~Mao, Q.~Wang, Z.~Wang, and S.~Chen, ``{Novel index modulation techniques: A
  survey},'' \emph{IEEE Commun. Surv. Tut.}, vol.~21, no.~1, pp. 315--348,
  2019.

\bibitem{10381617}
T.~Mao, Z.~Zhou, Z.~Xiao, C.~Han, and Z.~Wang, ``{Index-modulation-aided
  terahertz communications with reconfigurable intelligent surface},''
  \emph{IEEE Trans. Wireless Commun.}, vol.~23, no.~7, pp. 8059--8070, Jul.
  2024.

\end{thebibliography}
\end{document}